\documentclass[12pt]{article}
\usepackage{amsmath}
\usepackage{multirow}
\usepackage{amsfonts}
\usepackage{amssymb,graphics,psfrag}
\usepackage{array,epsfig,multirow,graphicx}
\usepackage{comment}
\usepackage{feynmp}

\usepackage[all,cmtip]{xy}

\numberwithin{equation}{section}
\numberwithin{table}{section}

\usepackage{hyperref}

\def\hybrid{\topmargin -20pt    \oddsidemargin 0pt
        \headheight 0pt \headsep 0pt
        \textwidth 6.25in      % A4 paper
        \textheight 9 in      % A4 paper
        \marginparwidth .875in
        \parskip 5pt plus 1pt
          \jot = 1.5ex
  }
\hybrid

\newcommand{\be}{\begin{equation}}
\newcommand{\ee}{\end{equation}} 
\newcommand{\bea}{\begin{eqnarray}}
\newcommand{\eea}{\end{eqnarray}}

\newcommand{\nn}{\nonumber}

\newcommand{\beq}{\begin{equation}}
\newcommand{\eeq}{\end{equation}}
\newcommand{\cS}{\mathcal{S}}

\newcommand{\cref}{{\bf [check ref]}}

\newcommand{\Ssu}{\cS_{\text{SU}(5)}}

\def\blfootnote{\xdef\@thefnmark{}\@footnotetext}
\long\def\symbolfootnote[#1]#2{\begingroup%
\def\thefootnote{\fnsymbol{footnote}}\footnote[#1]{#2}\endgroup}

\begin{document}

\baselineskip=15pt

\begin{titlepage}
\begin{flushright}
\parbox[t]{1.08in}{UPR-1253-T}
\end{flushright}

\begin{center}

\vspace*{ 1.2cm}

{\Large \bf \phantom{.}\text{\hspace{-0.5cm}F-Theory Compactifications with Multiple U(1)-Factors}: Addendum}

\vskip 1.2cm

\renewcommand{\thefootnote}{}
\begin{center}
 {Mirjam Cveti\v{c}$^{1,2}$,  Denis Klevers$^1$  and Hernan Piragua$^1$}
\end{center}
\vskip .2cm
\renewcommand{\thefootnote}{\arabic{footnote}} 

$\,^1$ {Department of Physics and Astronomy,\\
University of Pennsylvania, Philadelphia, PA 19104-6396, USA} \\[.3cm]

$\,^2$ {Center for Applied Mathematics and Theoretical Physics,\\ University of Maribor, Maribor, Slovenia}\\[.3cm]

{cvetic\ \textsf{at}\ cvetic.hep.upenn.edu, klevers\ \textsf{at}\ sas.upenn.edu, hpiragua\ \textsf{at}\ sas.upenn.edu } 

 \vspace*{0.8cm}

\end{center}

\vskip 0.2cm
 
\begin{center} {\bf ABSTRACT } \end{center}

The purpose of this note is to extend the results obtained in 
\cite{Cvetic:2013nia}  in two ways. First, the six-dimensional F-theory 
compactifications with U(1)$\times$U(1) gauge symmetry  on elliptic  
Calabi-Yau  threefolds,  constructed  as a  hypersurface  in   $dP_2$ 
fibered over the base $B=\mathbb{P}^2$  \cite{Cvetic:2013nia} , are 
generalized to Calabi-Yau threefolds elliptically fibered over an 
arbitrary two-dimensional base  $B$. While the representations of the 
matter hypermultiplets remain unchanged, their multiplicities are 
calculated for an arbitrary $B$.  Second, for a  specific non-generic 
subset of such Calabi-Yau threefolds   we engineer   
SU(5)$\times$U(1)$\times$U(1)  gauge symmetry. We summarize the 
hypermultiplet matter representations,  which remain the same as for the 
choice of the base $B=\mathbb{P}^2$  \cite{Cvetic:2013uta}, and determine 
their multiplicities for an arbitrary $B$. We also verify that the 
obtained spectra cancel anomalies both for  U(1)$\times$U(1) and 
SU(5)$\times$U(1)$\times$U(1).

\hfill {July, 2013}
\end{titlepage}

\tableofcontents

%\addtolength\topmargin{50pt}
%\addtolength\textheight{-105pt}

%
%%%%%%%%%%%%%%%%%%%%%%%%%%%%%%%%%%%%%%%%%%%%%%%%%%%%%%%%%%%%%%%%%%%%%%%%%%%%%%%%%%%%%%%%%%%%%%%%%%
%%%%%%%%%%%%%%%%%%%%%%%%%%%%%%%%%%%%%%%%%%%%%%%%%%%%%%%%%%%%%%%%%%%%%%%%%%%%%%%%%%%%%%%%%%%%%%%%%%

\section{Introduction}
In this Addendum  we extend the results of  \cite{Cvetic:2013nia}   to  
constructions  of elliptically fibered Calabi Yau threefolds $\hat{X}$ 
with Mordell-Weil group of rank two for an arbitrary  two-dimensional 
base $B$,  leading to a most general construction of six-dimensional 
F-theory compactifications with $\text{U}(1)\times \text{U}(1)$ gauge 
symmetry. While the explicit examples  in \cite{Cvetic:2013nia} were 
given for the base  $B=\mathbb{P}^2$, a generalization to an arbitrary 
base  turns out to depend only on the information about 
the three divisors in the base: the anti-canonical divisor of $B$ and two 
more divisors  $\cS_9$ and $\cS_7$ that are projections with respect to  
$\pi:\,\hat{X}\rightarrow B$ to the 
homology of the base $B$ of particular intersections of the divisor 
classes $S_Q$, $S_R$ and $S_P$ of the rational sections 
${\hat s}_Q $,  ${\hat s}_R$ and the zero section ${\hat s}_P$, namely  
$\cS_9=\pi(S_P \cdotp S_R)$ and $\cS_7=\pi(S_Q \cdotp S_R)$.

The hypermultiplet matter  representations appear at codimension two 
singularities where the fiber degenerates in a way that is independent of 
the base $B$.  Therefore their representations under $\text{U}(1)\times \text{U}(1)$ 
gauge symmetry, which  were determined  in \cite{Cvetic:2013nia}, remain 
the same for an arbitrary base $B$.  On the other hand,  their 
multiplicities do depend on the base choice, and in this Addendum we 
extend the calculation of  hypermultiplet matter multiplicities to an 
arbitrary base and  verify anomaly cancellation. These results therefore 
determine hypermultiplet matter representations and their multiplicities 
for a general six-dimensional compactification of F-theory with 
$\text{U}(1)\times \text{U}(1)$ gauge symmetry.

Furthermore, in the original paper we gave a specific  toric example of a 
non-generic elliptic Calabi Yau  threefold with the gauge group 
$\text{SU}(5)\times \text{U}(1)\times \text{U}(1)$. The purpose was to 
show that it was possible to add a non-Abelian group on divisors in the 
base, along with  the Abelian sector $\text{U}(1)\times \text{U}(1)$.  
The hypermultiplet 
matter representations, which again depend only on the codimension two 
singularity structure of the fiber, were subsequently given in 
\cite{Cvetic:2013uta}.   Here we engineer  
$\text{SU}(5)\times \text{U}(1) \times \text{U}(1)$ 
gauge symmetry on a non-generic elliptically fibered Calabi-Yau threefold 
over an arbitrary base $B$.  We derive the multiplicities of the 
hypermultiplet  matter representations  and verify  anomaly cancellation.  
In this analysis only the information on one additional vertical divisor 
$\cS_{\text{SU(5)}}$, which is associated with the $\text{SU}(5)$ gauge 
symmetry, is needed.

We follow closely the notation of the  subsequent paper  
\cite{Cvetic:2013uta} where four-dimensional F-theory compactifications 
with $\text{U}(1)\times \text{U}(1)$ and  
$\text{SU}(5)\times\text{U}(1)\times \text{U}(1)$ gauge symmetry were 
studied. We have also taken some formulae, tables and figures from this 
paper.

For related work on  F-theory compactifications on Calabi-Yau manifolds 
with rank two Mordell-Weil group, see
\cite{Borchmann:2013jwa,Grimm:2013oga,Braun:2013nqa,Borchmann:2013hta}.

The Addendum is structured in the following way. In Section 2 we  
systematically construct  Calabi-Yau threefolds  with 
$\text{U}(1)\times \text{U}(1)$ 
gauge symmetry,  which are elliptically fibered over an arbitrary base, 
calculate the multiplicities of all hypermultiplet  matter 
representations and verify  anomaly cancellation. In Section 3 we 
summarize the spectrum for a specific non-generic fibration with 
$\text{SU}(5)\times \text{U}(1)\times \text{U}(1) $ gauge symmetry,  and 
calculate multiplicities for Calabi-Yau threefolds with an arbitrary 
base. We also show that these spectra cancel six-dimensional anomalies.

\section{Compactifications for a General Base: U(1)$\times$U(1)}

In this Section we work out the spectrum of a general F-theory
compactification to six dimensions on elliptically fibered 
Calabi-Yau threefolds  $\pi:\,\hat{X}\rightarrow B$ realizing a rank two 
Abelian sector. Consistency of
the theory is shown at the level of  anomaly cancellation.

\subsection{Construction of the  Fibration}

The general elliptic fiber with a rank two Mordell-Weil group 
is given by the generic elliptic curve $\mathcal{E}$ in $dP_2$
\cite{Cvetic:2013nia,Borchmann:2013jwa}, which takes 
the following form:
\beq \label{eq:CYindP2}
	p\!=\! u (s_1u^2e_1^2e_2^2 + s_2 u v e_1 e_2^2 + s_3  v^2e_2^2 
	+ s_5 u w e_1^2e_2+ s_6 vwe_1e_2 + s_8 w^2 e_1^2)
	+ s_7 v^2 we_2 + s_9 v w^2e_1\,,
\eeq
where $[u:v:w:e_1:e_2]$ are the homogeneous coordinates on $dP_2$ and
the $s_i$'s are numbers in $\mathbb{C}$.

The elliptic fibrations of this curve over the base $B=\mathbb{P}^2$
have been constructed in \cite{Cvetic:2013nia}. In this Section,
we extend the construction to an arbitrary base $B$. 
Here we closely follow the derivation and the notation of 
\cite{Cvetic:2013uta}.  
These elliptic threefolds $\hat{X}$ can be described as a hypersurface in 
a four-dimensional ambient space.

We begin by constructing this ambient space. It is given by a 
$dP_2$-fibration over the two-dimensional base $B$,
\beq \label{eq:dP2fibration}
	\xymatrix{
	dP_2 \ar[r] & 	dP^B_2(\cS_7,\cS_9) \ar[d]\\
	& B\,
	}
\eeq
In this fibration the homogeneous coordinates of $dP_2$ are lifted to 
sections of line bundles over the base $B$. Using the three
$\mathbb{C}^*$-actions on $dP_2$ three of these coordinates can be chosen 
to transform in the trivial line bundle on the base without loss of 
generality. We can choose the following  assignments of line bundles for
the remaining two homogeneous coordinates:
\beq \label{eq:LBassignment}
	u\in \mathcal{O}_B(\cS_9+ [K_{B}])\,,\qquad v\in \mathcal{O}_B(\cS_9-\cS_7)\,,
\eeq
where $[K_{B}]$ is the canonical bundle on $B$ and $\cS_7,\,\cS_9$ are 
arbitrary divisors on $B$. We note that \eqref{eq:dP2fibration} is the 
natural generalization of eq. (3.30) in \cite{Cvetic:2013nia}.

Next, we construct an elliptically fibered Calabi-Yau threefold with 
its general elliptic fiber in $dP_2$. It is described as the 
hypersurface \eqref{eq:CYindP2} in the ambient space 
\eqref{eq:dP2fibration}. The Calabi-Yau condition enforces that the 
coefficients $s_i$ are lifted to sections of appropriate line bundles
on $B$. To see this, we calculate the total Chern class of the 
ambient space via adjunction,
\beq \label{eq:c(dP2B)}
	c(dP_2^B)=
	c(B) (1 + H - E_1 - E_2 + \cS_9 -[K_B^{-1}]) (1 + H - E_2 + \cS_9 - \cS_7) (1 +  H - E_1) 
	(1 +  E_1) (1 + E_2)\,,
\eeq
where $H$, $E_1$ and $E_2$ are the three divisor classes on $dP_2$, 
c.f., eq. (3.24) in \cite{Cvetic:2013nia} as well as  eq. (2.2) of 
\cite{Cvetic:2013uta}.
This yields the following anti-canonical bundle:
\beq \label{eq:antiKdP2}
	K_{dP^B_2}^{-1}=\mathcal{O}(3H-E_1-E_2+2\cS_9-\cS_7)\,,
\eeq
where for brevity of notation we suppressed the dependence on $\cS_7$, $\cS_9$ on the left side 
of (\ref{eq:c(dP2B)}) and (\ref{eq:antiKdP2}). Requiring that the hypersurface \eqref{eq:CYindP2}
is a section of the anti-canonical bundle, we obtain the following
assignment of line bundles for the coefficients $s_i$:
\beq \label{eq:sectionsFibration}
\text{
\begin{tabular}{c|c}
\text{Section} & \text{Bundle}\\
\hline
	$u$&$\mathcal{O}(H-E_1-E_2+\cS_9+[K_B])$\rule{0pt}{13pt} \\
	$v$&$\mathcal{O}(H-E_2+\cS_9-\cS_7)$\rule{0pt}{12pt} \\
	$w$&$\mathcal{O}(H-E_1)$\rule{0pt}{12pt} \\
	$e_1$&$\mathcal{O}(E_1)$\rule{0pt}{12pt} \\
	$e_2$&$\mathcal{O}(E_2)$\rule{0pt}{12pt} \vspace{1.73cm}\\
\end{tabular}
}\qquad \text{
\begin{tabular}{c|c}
\text{Section} & \text{Bundle}\\
\hline
	$s_1$&$\mathcal{O}(3[K_B^{-1}]-\cS_7-\cS_9)$\rule{0pt}{13pt} \\
	$s_2$&$\mathcal{O}(2[K_B^{-1}]-\cS_9)$\rule{0pt}{12pt} \\
	$s_3$&$\mathcal{O}([K_B^{-1}]+\cS_7-\cS_9)$\rule{0pt}{12pt} \\
	$s_5$&$\mathcal{O}(2[K_B^{-1}]-\cS_7)$\rule{0pt}{12pt} \\
	$s_6$&$\mathcal{O}([K_B^{-1}])$\rule{0pt}{12pt} \\
	$s_7$&$\mathcal{O}(\cS_7)$\rule{0pt}{12pt} \\
	$s_8$&$\mathcal{O}([K_B^{-1}]+\cS_9-\cS_7)$\rule{0pt}{12pt} \\
	$s_9$&$\mathcal{O}(\cS_9)$ \rule{0pt}{12pt} 
\end{tabular}
}
\eeq
Here, in the first column,  we have also summarized the line bundles of 
the homogeneous coordinates on $dP_2$.

The rational sections $\hat{s}_P$, $\hat{s}_Q$ and 
$\hat{s}_R$ are given by $e_2=0$, $e_1=0$ and $u=0$ in $\hat{X}$. From 
this it follows that their classes, denoted by capital letters, are given 
by:
\beq \label{eq:sections}
S_P = E_2\,, \qquad S_Q = E_1\,, \qquad S_R = H - E_1 - E_2 + \cS_9 + [K_B]\,,
\eeq
as in \cite{Cvetic:2013nia}. We choose $\hat{s}_P$ as the zero section of 
the elliptic fibration of $\hat{X}$.

From the assignments \eqref{eq:sectionsFibration} we see that the
divisors $\cS_7$, $\cS_9$ are precisely the vanishing loci of the
sections $s_7$, $s_9$ in \eqref{eq:dP2fibration}, which is the reason
for the choice of their labels. As mentioned before, these two divisors
geometrically encode the intersections of the rational sections. Indeed, 
we observe  the following intersections:
\beq \label{eq:defS9S7}
\cS_7=\pi(S_Q \cdotp S_R)\,, \qquad \cS_9=\pi(S_P \cdotp S_R)\,, \qquad 0=\pi(S_P \cdotp S_Q)\,,
\eeq
where $\pi$ denotes the projection to the (co-)homology of the base $B$.
 
The assignment of sections in \cite{Cvetic:2013nia} for the base 
$B=\mathbb{P}^2$ can be recovered from \eqref{eq:sectionsFibration} by 
making the replacements $[K_B^{-1}]\rightarrow 3 H_B$, 
$\cS_9\rightarrow n_2 H_B$ and $\cS_7\rightarrow n_{12} H_B$, where $H_B$ 
is the hyperplane class of the base.

\subsection{Hypermultiplet Matter Representations and Multiplicities}

As mentioned in the Introduction, locally the fiber degenerates in the 
same fashion at codimension two  as for the case $B=\mathbb{P}^2$. 
Furthermore, the intersections of the rational sections 
with the components of the reducible fiber do not depend on the global 
geometry of $B$. This 
implies that the $U(1)$-charges of matter fields are independent of the 
base $B$, as can be seen from  the charge formula
\beq
	q_m=(S_m-S_P)\cdot c_{\text{mat}}\, , 
\eeq
where $c_{\text{mat}}$ is an isolated $\mathbb{P}^1$ in the fiber 
corresponding to a matter field and where we collectively labeled
the  divisor classes $S_m=(S_Q,S_R)_m$, with $m=1,2$. 

Furthermore, the functional dependencies of 
the polynomials describing the codimension two loci supporting matter on 
the sections $s_i$ are base-independent.  
However, the
multiplicities of matter do depend on 
$B$ because the line bundles in which the $s_i$'s take values are base 
dependent, see \eqref{eq:sectionsFibration}. Nevertheless, in the 
following we can still calculate the matter multiplicities directly for 
an arbitrary base by applying the  results in Section 4 of 
\cite{Cvetic:2013nia} and  the intersection theory on $B$.

First, we calculate the multiplicities of the singlets in Section 4.3 of
\cite{Cvetic:2013nia}.  These arise at  codimension two singularities of 
the fibration due to the rational sections being ill-defined. These 
codimension two loci are complete intersections in the base $B$ and their 
multiplicities are the number of intersection points given by
\beq \label{eq:multCharge2}
\text{
 \centering
 \begin{tabular}{|c|c|c|} \hline 
 Loci 		& Representation 	& Multiplicity \rule{0pt}{13pt}\\ \hline
 $s_7=s_3=0$	&$\mathbf{1}_{(-1,1)}$	  		& $%x_4=
\cS_7\cdot ([K_B^{-1}]+\cS_7-\cS_9)$\rule{0pt}{12pt}	\\ \hline
 $s_7=s_9=0$	& $\mathbf{1}_{(0,2)}$	  		& $%x_5=
 \cS_7\cdot\cS_9$	\\ \hline
 $s_9=s_8=0$	&$\mathbf{1}_{(-1,-2)}$	  		& $%x_6=
\cS_9\cdot ([K_B^{-1}]+\cS_9-\cS_7)$\rule{0pt}{1Em}\\ \hline
 \end{tabular}
}
\eeq
where we used \eqref{eq:sectionsFibration}. Here $\cdot$ denotes the 
intersection product on $B$. 

Second, we calculate the multiplicities of the singlets in
Sections 4.1 and 4.2 of \cite{Cvetic:2013nia}. These arise at  codimension two singularities
 where a rational section collides with  a singularity in the Weierstrass fibration. 
We first calculate the 
line bundles of the polynomials of the complete intersections that 
contain the respective codimension two loci. They read
\beq \label{eq:codim2loci2}
\text{
\begin{tabular}{|c|c||c|c|} \hline
Polynomial & Bundle & Polynomial & Bundle \\ \hline
$\delta g_6^\prime$ & $\mathcal{O}(2[K_B^{-1}]+2\cS_9-\cS_7$) & $g_9^\prime$ & $\mathcal{O}(3[K_B^{-1}]+\cS_9-\cS_7)$  \rule{0pt}{1Em}\\ \hline
$g^Q_9$  & $\mathcal{O}(3[K_B^{-1}])$ & $\hat{g}^Q_{12}$ & $\mathcal{O}(4[K_B^{-1}])$\rule{0pt}{1Em}\\\hline
$g^R_{9+3n_2}$  & $\mathcal{O}(3[K_B^{-1}]+3\cS_9)$ & $\hat{g}^R_{12+4n_2}$ & $\mathcal{O}(4[K_B^{-1}]+4\cS_9)$\rule{0pt}{1Em} \\ \hline
\end{tabular} 
}
\eeq
The polynomials $\delta g_6^\prime$ and $g_9^\prime$ were defined in eq. 
(4.19) of \cite{Cvetic:2013nia}. Their zero loci contain the loci of the 
representation $\mathbf{1}_{(1,1)}$. Similarly, $g^Q_9$ and 
$\hat{g}^Q_{12}$ were defined  in eqs. (4.6) and (4.7) of 
\cite{Cvetic:2013nia}, respectively. Their zero loci contain the loci of 
the representation  $\mathbf{1}_{(1,0)}$. Finally,
the loci of $\mathbf{1}_{(0,1)}$ are contained in the zero loci of 
$g^R_{9+3n_2}$ and $\hat{g}^R_{12+4n_2}$ defined in  eqs. (4.10) and 
(4.11) of \cite{Cvetic:2013nia}, respectively.

The polynomials \eqref{eq:codim2loci2} also have zeros  at the loci 
corresponding to the fields  in (\ref{eq:multCharge2}) that have already 
been counted.
We have to subtract these zeros with the appropriate order in order to 
obtain the multiplicities of the  matter in Sections 4.1 and 4.2 of
\cite{Cvetic:2013nia}. Because the functional dependencies of 
the polynomials in \eqref{eq:codim2loci2} on the $s_i$'s  are the same as 
for the case $B=\mathbb{P}^2$, the 
order of the zeroes determined in Section 4.4  of 
\cite{Cvetic:2013nia}, using resultant techniques, still holds. Invoking 
these results, the rest of the 
multiplicities are
\beq \label{eq:multCharge1}
\text{
\begin{tabular}{|c|c|} \hline
Representation & Multiplicity \\ \hline
$\mathbf{1}_{(1,0)}$ & $6 [K_B^{-1}]^2 + [K_B^{-1}] \cdot (4\cS_7 -5\cS_9)- 2 \cS_7^2 + \cS_7 \cdot \cS_9 + \cS_9^2$\rule{0pt}{1.1Em}\\ \hline
$\mathbf{1}_{(0,1)}$ & $6 [K_B^{-1}]^2 + [K_B^{-1}]\cdot  (4\cS_7+4\cS_9) - 2 \cS_7^2  - 2 \cS_9^2$\rule{0pt}{1Em}\\ \hline
$\mathbf{1}_{(1,1)}$ & $6 [K_B^{-1}]^2 + [K_B^{-1}] \cdot(-5 \cS_7+4\cS_9) + \cS_7^2  + \cS_7 \cdot \cS_9 - 2 \cS_9^2$\rule{0pt}{1Em} \\ \hline
\end{tabular}
}
\eeq

We note that the results in Section 4 of \cite{Cvetic:2013nia}  for the 
special base  $B=\mathbb{P}^2$ are recovered by making the 
identifications $[K_B^{-1}]\rightarrow 3 H_B$, 
$\cS_7 \rightarrow n_{12} H_B$, $\cS_9 \rightarrow n_{2} H_B$ and 
$H_B^2\rightarrow 1$, where again $H_B$ is the hyperplane class of the 
base $B$.

\subsection{Anomaly Cancellation}

With the  hypermultiplet representations and their multiplicities at 
hand, six-dimensional anomaly cancellation can be verified. For the 
convenience of the reader we summarize below the expressions for the 
respective mixed Abelian-gravitational, pure Abelian and pure 
gravitational anomaly cancellation:\footnote{For further details,  see 
Section 5 of \cite{Cvetic:2013nia}.}

\bea \label{eq:6dAnomalies}
 	K_B\cdotp b_{mn} &= &-\textstyle{\frac{1}{6}}\sum_{\underline{q}} x_{q_m, q_n} q_m q_n\,,\nn\\
 	b_{mn} \cdotp b_{kl} + b_{mk} \cdotp b_{nl} +b_{ml} \cdotp b_{nk}  &=& \sum_{\underline{q}} x_{q_m,q_n,q_k,q_l} q_m q_n q_k q_l\,,\nn\\
 	273=H-V+ 29T\,,&&\,\, K_B\cdot K_B=9-T\,.
\eea
Here $x_{q_m,q_n}$ and  $x_{q_m,q_n,q_k,q_l}$ denote the
number of matter hypermultiplets with charges $(q_m,q_n)$ and 
$(q_m,q_n,q_k,q_l)$ under $\text{U}(1)_m\times \text{U}(1)_n$, respectively,
$\text{U}(1)_m\times \text{U}(1)_n\times\text{U}(1)_k\times \text{U}(1)_l$. In addition, $H$, $V$ and $T$ denote the total 
number of hyper-, vector- and tensor-multiplets, respectively. 
The $b_{mn}$ denote curves in the base $B$ defined as
\bea \label{eq:b_mnAbelian}
 b_{mn} = -\pi( \sigma(\hat{s}_m) \cdotp \sigma(\hat{s}_n) ) =\left( \begin{array}{cc}
                    -2[K_B] & \cS_9-\cS_7  - [K_B] \\
                    \cS_9-\cS_7  - [K_B] & 2(\cS_9-[K_B])
                   \end{array} \right)_{mn}\,.
\eea
 Here we have collectively denoted  rational sections 
$\hat{s}_m=(\hat{s}_Q,\hat{s}_R)_m$ and their divisor classes  
$S_m=(S_Q,S_R)_m$,  with $m=1,2$ as before.
The Shioda map of rational sections ${\hat s}_m$ to  corresponding 
divisor classes $\sigma({\hat s}_m)$ gives:
 \beq\label{eq:AbelianShoida}
 \sigma({\hat s}_Q)=S_Q-S_P-[K_B^{-1}]\, , \quad \sigma({\hat s}_R)=S_R-S_P-[K_B^{-1}]-\cS_9\, .
 \eeq
See Section 2 of \cite{Cvetic:2013nia} as well as Section 2 of 
\cite{Cvetic:2013uta} for further details. In (\ref{eq:b_mnAbelian})  we 
have further used  \eqref{eq:defS9S7}. 

Inserting \eqref{eq:b_mnAbelian} as well as the multiplicities 
\eqref{eq:multCharge2} and \eqref{eq:multCharge1} into the anomaly 
cancellation
equations \eqref{eq:6dAnomalies}, we see that all equations are satisfied 
and thus all anomalies are 
cancelled. This verifies the six-dimensional anomaly cancellation for the 
general F-theory compactification with U$(1)\times$U(1) gauge symmetry 
over an arbitrary base $B$.\footnote{The base $B$ has to admit a generic
elliptic fibration of the form \eqref{eq:CYindP2}, i.e.~all line bundles 
in \eqref{eq:sectionsFibration} have to have generic sections $s_i$. See 
the discussion in  Section 2 of \cite{Cvetic:2013uta}.} 

\section{Compactifications for a General Base:
SU(5)$\times$U(1)$^2$}

In Section 6.3 of \cite{Cvetic:2013nia}  an example with 
SU(5)$\times$U(1)$\times$U(1) gauge symmetry was constructed as a 
non-generic elliptic Calabi-Yau threefold with $B=\mathbb{P}^2$. It
was described as a toric hypersurface in a four-dimensional toric ambient
space obtained by  a toric resolution corresponding to one particular 
SU(5)-Top.   
In this Section we generalize this construction to an elliptically 
fibered Calabi-Yau threefold with an arbitrary base $B$. 

In order to 
align the notation with the one of \cite{Cvetic:2013uta},  we  
interchange the variables $d_2$ and $d_4$ in \cite{Cvetic:2013nia} and 
their respective divisor classes. After this change, the nodes of the
affine Dynkin diagram of SU(5) correspond to $d_0,\, \ldots,d_4$ in 
cyclic order starting with the extended node, see the left figure in  
Figure \ref{fig:SU5Figures}.

For a specific choice \cite{Cvetic:2013nia} of a non-generic elliptic  fibrations, where the following sections $s_i$ are chosen 
to vanish at $z_2=0$ in the following way:
\bea \label{eq:s'SU5}
s_1 = z_2^3 s'_1\,, \qquad  s_2 = z_2^2 s'_2\,, \qquad s_3 = z_2^2 s'_3\,, \qquad s_5 = z_2 s'_5\,,
\eea 
there is an $A_4$ singularity on 
top of the hyperplane divisor $\cS_{\text{SU(5)}}^b=H_B$ in 
$B=\mathbb{P}^2$, thus resulting in  the SU(5) gauge symmetry at $z_2=0$.   This  was confirmed by computing the discriminant $\Delta$ of the Weierstrass fibration, which at $z_2=0$ takes  the 
form
\be \label{eq:DeltaSU5}
 \Delta = -z_2^5  \left(\beta_5^4 P + z_2 \beta_5^2P_2 R
  + \mathcal{O}(z_2^2)\right)\, ,
\ee
where 
\be \label{eq:betaP}
	\beta_5=s_6\,, \quad P: = P_1 P_2 P_3 P_4 P_5 =  (s'_2 s'_5 - s'_1 s_6) s_7 (s'_2 s_7-s'_3 s_6) s_8 (s_6 s_9-s_7 s_8 )\,,
\ee
and $R$ is a polynomial in $s_i$'s with no common factors. We refer to the discussion around eq. (6.17) of \cite{Cvetic:2013nia} for
further details.

The $A_4$ singularity is resolved by four blow-ups in the four-
dimensional ambient space \eqref{eq:dP2fibration} with $B=\mathbb{P}^2$:
\beq \label{eq:resMapSU5}
\pi_{\text{SU(5)}}\,:\quad w= d_1 d_2^2 d_3^3 d_4^2 \tilde{w}\,, \quad v= 
d_1 d_2 d_3 d_4  \tilde{v}\,, \quad u =d_3 d_4  \tilde{u}\,, \quad z_2 = d_1 d_2 d_3 d_4  \tilde{z}\, .
\eeq
Here we introduced the sections $d_i$ of the divisor bundles 
$\mathcal{O}(D_i)$ associated with the exceptional divisors 
$D_i:=\{d_i=0\}$ of the resolution. The Stanley-Reissner ideal of
this resolution reads
\bea \label{eq:SR-SU5}
SR&=&     \pi_{\text{SU}(5)}^*(SR_B)\cup \{ v e_1, e_1 e_2, u v, u w, w e_2 \}\cup \{ z_2 d_3, d_1 d_3, d_1 d_4\} \cup 
\{ d_1 e_1, z_2 e_1,\nn \\  &&d_2e_1, d_4 e_1,d_1 u,   d_2 u, 
d_1 e_2, d_2 e_2, d_3 e_2, d_4 e_2,  z_2 w, d_1 w, d_4 w, d_3 v, d_4 v,  z_2 d_2 v \}\, ,
%\cup  \pi_{\text{SU}(5)}^*(SR_B)
\eea
where we pulled back the Stanley-Reissner ideal $SR_B$ of the base
using \eqref{eq:resMapSU5}.

Along the same lines as in the previous Section it is possible to 
generalize this example to elliptic Calabi-Yau threefolds 
$\hat{X}_{\text{SU}(5)}$ over an arbitrary base. 
The threefold $\hat{X}_{\text{SU}(5)}$ is still described as a 
hypersurface in the blow-up of \eqref{eq:dP2fibration} via 
\eqref{eq:resMapSU5} for a general $B$. We denote this new ambient space 
as $\widehat{dP}_2^B(\cS_7,\cS_9)$. The divisor 
$\cS_{\text{SU}(5)}^b$, which  supports  the $A_4$ singularity, is replaced 
by a general divisor in the base 
$B$ associated with a section $z$,  replacing the section $z_2$ above. 

In the elliptic Calabi-Yau threefold $\hat{X}_{\text{SU}(5)}$ over a 
general base $B$, the sections $s_i'$ ($i=1,2,3,5$) are promoted to sections of the following line 
bundles:
\beq \label{eq:sectionsSU50}
\text{
\begin{tabular}{c|c}
\text{Section} & \text{Bundle}\\
\hline
$s'_1$&$\mathcal{O}(3[K_B^{-1}]-\cS_7-\cS_9-3\Ssu)$\rule{0pt}{13pt} \\
	$s'_2$&$\mathcal{O}(2[K_B^{-1}]-\cS_9-2\Ssu)$\rule{0pt}{12pt} \\
	$s'_3$&$\mathcal{O}([K_B^{-1}]+\cS_7-\cS_9-2\Ssu)$\rule{0pt}{12pt} \\
	$s'_5$&$\mathcal{O}(2[K_B^{-1}]-\cS_7-\Ssu)$\rule{0pt}{12pt} \\
\end{tabular}
}
\eeq
by the Calabi-Yau condition.
Here we introduced the vertical divisor 
$\cS_{\text{SU}(5)}=\pi^*(\cS_{\text{SU}(5)}^b)$ in 
$\hat{X}_{\text{SU}(5)}$.
This can be seen by first performing the proper transform of 
\eqref{eq:CYindP2} under \eqref{eq:resMapSU5} using \eqref{eq:s'SU5}
and  then computing the anti-canonical bundle of the  blown-up ambient 
space \eqref{eq:dP2fibration}, which reads
\beq \label{eq:KdP2SU5}
K_{\widehat{dP}_2^B}^{-1} = \pi_{\text{SU}(5)}^*\big(K^{-1}_{dP_2^B}\big) 
- 2D_1-3D_2-5 D_3-4 D_4\,.
\eeq
Here $\pi_{\text{SU}(5)}^*$ is the pullback (\ref{eq:resMapSU5})
and we refer also to Appendix C of \cite{Cvetic:2013uta} for more details
on the total Chern class of $\widehat{dP}_2^B(\cS_7,\cS_9)$.
Requiring $\hat{X}_{\text{SU}(5)}$ to be a section of 
$K_{\widehat{dP}_2^B}^{-1}$ we immediately obtain the assignments
\eqref{eq:sectionsSU50}.

\subsection{Hypermultiplet Matter Representations}

As pointed out in Sections 2 the hypermultiplet matter representations do 
not depend on the choice of the  base, and  for the non-generic elliptic 
fibration  (\ref{eq:s'SU5}) the representations of  $\text{SU(5)}\times 
\text{U}(1)\times \text{U}(1)$ were subsequently  determined in 
\cite{Cvetic:2013uta}.  In this Subsection we therefore present only a 
summary of the key results and  refer the reader to Section 6 of 
\cite{Cvetic:2013uta} for the details of these calculations. 

\begin{figure}[ht!]
\centering
  \includegraphics[scale=0.5]{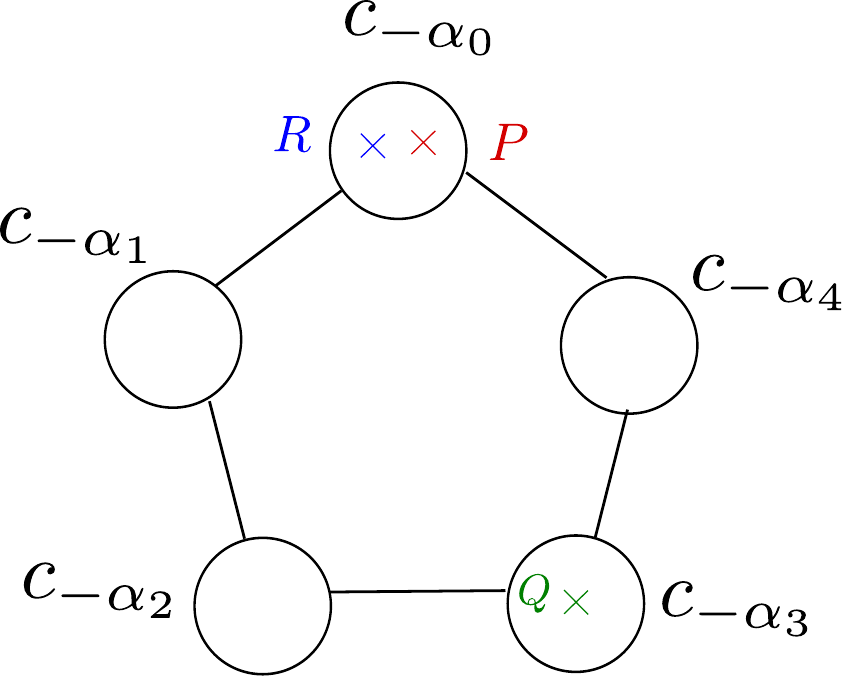} \qquad
  \includegraphics[scale=0.5]{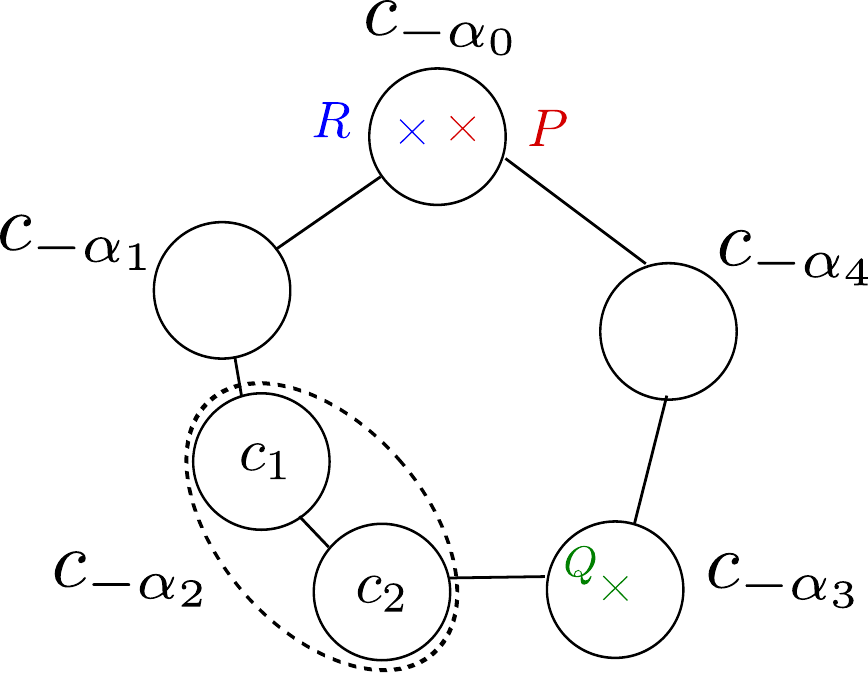} \qquad
  \includegraphics[scale=0.5]{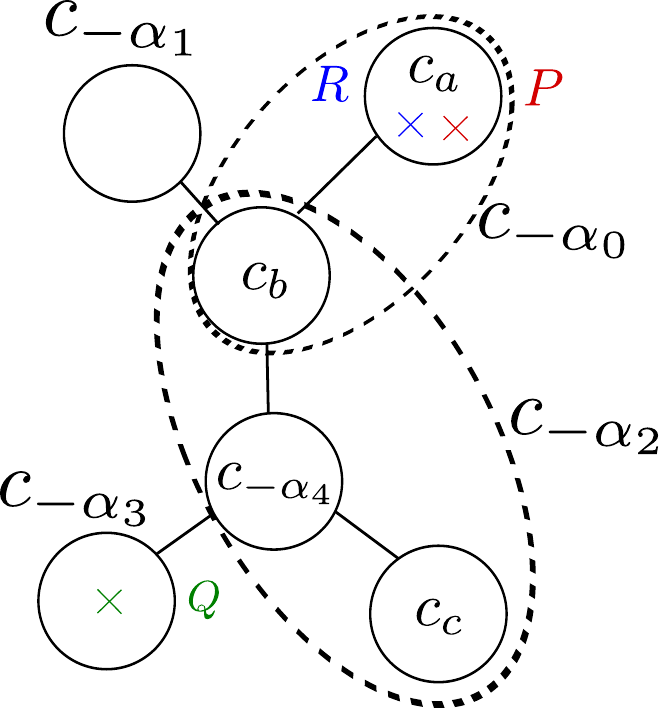}
  \caption{Left: $I_4$ singularity over a generic point of the divisor 
  $z=0$, marked according to the interaction of the nodes with 
  $\{S_P,S_Q,S_R\}$.   Center: $A_5$ singularity with  local SU(6) enhancement 
  at  $z=P_1=0$. Right: $D_5$ singularity with  local SO(10)  enhancement at 
  $z=\beta_5=0$. The dashed lines indicate  nodes of the Dynkin
diagram of SU(5) that split at a codimension two locus under 
consideration.}
  \label{fig:SU5Figures}
\end{figure}

We first summarize the intersection numbers on $\hat{X}_{\text{SU}(5)}$,
which  can be calculated using the representation of the even-dimensional 
cohomology ring $H^{(*,*)}(\hat{X}_{\text{SU}(5)})$ as the quotient 
polynomial ring in the divisors $\{ S_P, S_Q, S_R, D_i,D_\alpha\}$ 
modulo the Stanley-Reissner ideal \eqref{eq:SR-SU5} with $z_2\rightarrow 
z$. 
With the full cohomology ring at hand the intersection 
matrix of the exceptional divisors over an arbitrary point of 
$\cS_{\text{SU(5)}}$ is determined as 
\beq 
%\label{eq:intsCartansSU5}
  D_I \cdotp D_J \cdotp D_\alpha  = -C_{IJ}  \cS_{\text{SU(5)}} \cdotp S_P \cdotp D_\alpha  \, , \ \ 
%\eeq
%\beq 
\label{eq:SU5Cartan}
(-C_{IJ}) = \left( \begin{array}{ccccc}
                 -2 &  1 &  0 &  0 &  1 \\
                 1  & -2 &  1 &  0 &  0 \\
                 0  & 1  & -2 &  1 &  0 \\
                 0  &  0 &  1 & -2 &  1 \\
                 1  &  0 &  0 &  1 & -2 
                \end{array}
                \right)\, , 
\eeq
where $I$ and $J$ are in the range $\{0,\dots, 4\}$  with  $D_0$  as the 
extended node,  and  $D_\alpha$ is an arbitrary vertical divisor. The 
divisors $d_i=0$, $i=1,\ldots, 4$, that are $\mathbb{P}^1$-fibrations 
over $\cS^b_{\text{SU(5)}}$ 
are the Cartan divisors of SU(5). The individual $\mathbb{P}^1$'s in 
the fiber are curves $c_{-\alpha_i}$ that  correspond to the simple roots 
$\alpha_i$ of SU(5), with $\alpha_0=-\sum_i\alpha_i$ the extended node; 
the curves intersect as  the affine Dynkin diagram of SU(5),  thus 
justifying the depiction by the left figure of Figure 
\ref{fig:SU5Figures}.
The Shioda map is of the following form:
\beq \label{eq:ShiodaMapSQSRSU5}
\sigma(\hat s_Q)=  S_Q - S_P - [K_B^{-1}]+\textstyle{\frac{1}{5}}(2 D_1+4 D_2+ 6 D_3+3 D_4)\,,\ \
\sigma(\hat s_R) =  S_R - S_P - [K_B^{-1}] -  \cS_9\,,
\eeq
which can also be shown using the presentation of 
$H^{(*,*)}(\hat{X}_{\text{SU}(5)})$ as a polynomial ring.

We now summarize  the hypermultiplet matter representations located at 
codimension two singularities of the fibration.  In addition to singlet 
representations of SU(5) summarized in the respective first columns in 
\eqref{eq:multCharge2} and \eqref{eq:multCharge1} there are also 
non-singlet representations.  The latter ones appear at
$z=P_i=0$ ($i=1,\ldots,5$)  and at
$z=\beta_5=0$, where $P_i$ and $\beta_5$ are defined in 
%\eqref{eq:DeltaSU5} and
 \eqref{eq:betaP}. 

At $z=P_i=0$ ($i=1,\ldots,5$)  the singularity enhances to an $A_5$ 
singularity with local SU(6) symmetry, where one of 
the $\mathbb{P}^1$'s at codimension one splits into a sum of two isolated 
$\mathbb{P}^1$'s. By calculating the 
SU(5)-weight of the split curve, the matter located 
at these loci can be seen to transform as the $\mathbf{5}$ representation 
of SU(5). As an example, see the center figure of  Figure 
\ref{fig:SU5Figures} where the Dynkin diagram of the fiber at $z=P_1=0$ 
is shown. 
The U(1)-charges of the $\mathbf{5}$ representations at each $P_i=0$
are calculated as the intersections of the isolated rational curves,
corresponding to the  weight one  of the $\mathbf{5}$ representation, and 
the Shioda map \eqref{eq:ShiodaMapSQSRSU5}. The results are summarized 
in the following Table:
\beq
\text{
\begin{tabular}{|c||c|c|c|c|c|} \hline
 Loci		& $z=P_1=0$ & $z=P_2=0$ & $z=P_3=0$ & $z=P_4=0$ & $z=P_5=0$ \rule{0pt}{13pt}\\ \hline 
 $(q_1, q_2)$ & $(-\tfrac{2}{5},0)$ & $(-\tfrac{2}{5},1)$ &$(\tfrac{3}{5},0)$ & $(\tfrac{3}{5},1)$ & $(-\tfrac{2}{5},-1)$ \rule{0pt}{13pt} \\ \hline
\end{tabular}
}
\eeq

At the single locus $z=\beta_5=0$, the singularity enhances to a $D_5$ 
singularity with local enhancement to SO(10) symmetry. 
The representation at this locus is $\mathbf{10}_{(1/5,0)}$. 
The splitting of the rational curves and their
intersections  over this codimension two locus is depicted on  the right figure 
 of Figure 
\ref{fig:SU5Figures}. 

\subsection{Hypermultiplet Matter Multiplicities}

In this Subsection we determine the multiplicities of all hypermultiplet 
representations for an arbitrary base $B$. 

We begin with the multiplicities of the SU(5)-singlets 
\eqref{eq:multCharge2}. We note that there is only a minor change due to 
the presence of the SU(5) gauge symmetry  since the line bundle of $s_3'$
is now given by \eqref{eq:sectionsSU50}. Thus, we immediately obtain: 
\beq \label{eq:multCharge2SU5}
\text{
 \centering
 \begin{tabular}{|c|c|c|} \hline 
 Loci		& Representation 	& Multiplicity \rule{0pt}{13pt}\\ \hline
 $s_7=s_3'=0$	&$\mathbf{1}_{(-1,1)}$	  		& $\cS_7\cdot([K_B^{-1}]+\cS_7-\cS_9-2\cS_{\text{SU(5)}})$\rule{0pt}{12pt}	\\ \hline
 $s_7=s_9=0$	& $\mathbf{1}_{(0,2)}$	  		& $\cS_7\cdot\cS_9$	\\ \hline
 $s_9=s_8=0$	&$\mathbf{1}_{(-1,-2)}$	  		& $\cS_9\cdot([K_B^{-1}]+\cS_9-\cS_7)$\rule{0pt}{1Em}\\ \hline
 \end{tabular}
}
\eeq

The multiplicities of  the  $\mathbf{5}$   representations and the $\mathbf{10}$ representation are 
calculated straightforwardly by counting the points in the loci 
$z=P_i=0$ ($i=1,\cdots,5$) and $z=\beta_5=0$, respectively.  Since these loci are complete 
intersections, the results read:
\beq \label{eq:mults5}
\text{\centering
\begin{tabular}{|c|c|c|} \hline
 Loci		& Representation 	& Multiplicity \rule{0pt}{13pt}\\ \hline 
 $z=P_1=0$&$\mathbf{5}_{(-2/5,0)}$ & 
$ \cS_{\text{SU(5)}}\cdotp(4[K_B^{-1}]-\cS_7-\cS_9-3\cS_{\text{SU(5)}})$\rule{0pt}{1.1Em}\\ \hline
$z=P_2=0$&  $\mathbf{5}_{(-2/5,1)}$ & $ \cS_{\text{SU(5)}}\cdotp\cS_7$ \\ \hline
$z=P_3=0$&$\mathbf{5}_{(3/5,0)}$ &$\cS_{\text{SU(5)}}\cdotp( 2[K_B^{-1}]-\cS_9+\cS_7-2\cS_{\text{SU(5)}})  $\rule{0pt}{1Em}\\ \hline
$z=P_4=0$& $\mathbf{5}_{(3/5,1)}$ &$\cS_{\text{SU(5)}}\cdotp([K_B^{-1}] + \cS_9 -\cS_7)$ \rule{0pt}{1Em}\\ \hline
$z=P_5=0$&  $\mathbf{5}_{(-2/5,-1)}$ &
$ \cS_{\text{SU(5)}}\cdotp ([K_B^{-1}]+\cS_9)$ \rule{0pt}{1Em}\\ \hline
$z=\beta_5=0$& $  \mathbf{10}_{(1/5,0)}$& $\cS_{\text{SU(5)}}\cdotp[K_B^{-1}]$\rule{0pt}{1Em} \\ \hline
\end{tabular}
}
\eeq
Here we have made use of \eqref{eq:sectionsFibration} and 
\eqref{eq:sectionsSU50}.

The multiplicities of the remaining singlets \eqref{eq:multCharge1} have 
to be calculated with special care. Similarly,  as in the pure 
U(1)$\times$U(1) case, some of the non-Abelian  representations are 
contained in the loci of the  singlets, i.e.~the loci of the non-Abelian 
matter are  roots of the polynomials in \eqref{eq:codim2loci2}. 
Thus, in order to avoid a 
double-counting, we have to subtract the  fields we have already
counted.
For instance, the multiplicity of the fields in the representation 
$\mathbf 1_{(1,1)}$ is obtained from the number of points in the complete 
intersection $\delta g_6'=g_9'=0$ after subtracting a fourth order zero, a third order zero and a double zero 
corresponding to the  representation $\mathbf 1_{(-1,-2)}$,  $\mathbf 5_{(3/5,1)}$ and 
$\mathbf 5_{(-2/5,1)}$, respectively.

The determination of the multiplicity for the representation $\mathbf 1_{(1,0)}$ is more 
involved because
the polynomial equations $g^Q_9=\hat{g}^Q_{12}=0$, which contain
this representation, vanish automatically along the GUT-divisor 
$\cS_{\text{SU}(5)}$ at $z=0$. Thus, in order to  determine its multiplicity we have to cancel global factors
of $z$ in both $g^Q_9=\hat{g}^Q_{12}=0$. Then we  can 
count the number of points in this complete 
intersection, where we  have to subtract zeroes of order one corresponding to  the
representations $\mathbf 1_{(1,1)}$, $\mathbf 1_{(-1,1)}$, $\mathbf 1_{(-1,-2)}$, $\mathbf 5_{(3/5,1)}$ and $\mathbf 5_{(3/5,0)}$.

Finally, in order to obtain the multiplicity of the representation 
$\mathbf 1_{(0,1)}$ we have to count the 
number of points in $g_6^R=g_9^R=0$ by subtracting the  zeros of order 
one,   sixteen, and  five  corresponding to  the representations   $\{ { \mathbf 1}_{(1,1)}, {\mathbf 1}_{(-1,1)} \} $,  $\{ {\mathbf 1}_{(0,2)},  {\mathbf 1} _{(-1,-2)} \}$,  and $\{ {\mathbf 5}_{(-2/5,1)}, {\mathbf 5}_{(3/5,-1)}, {\mathbf 5}_{(-2/5,-1)} \}$, respectively.

We summarize the results of this analysis in the following Table:
\beq \label{eq:multcharge1SU5}
\text{
\begin{tabular}{|c|c|} \hline
Representation \!\!&\!\! Multiplicity \\ \hline
$\mathbf{1}_{(1,0)}$ \!\!&\!\! $6 [K_B^{-1}]^2 + [K_B^{-1}] \cdot(-12 \cS_{\text{SU(5)}} + 4 \cS_7 - 5 \cS_9)$\rule{0pt}{1.2Em}\\ 
 \!\!&\!\!  $+ 6 \cS_{\text{SU(5)}}^2 + \cS_{\text{SU(5)}}\cdot(- \cS_7+5\cS_9) - 2 \cS_7^2 
 + \cS_7 \cdot\cS_9 + \cS_9^2$\\ \hline
$\mathbf{1}_{(0,1)}$ \!\!&\!\! $6 [K_B^{-1}]^2 + [K_B^{-1}] \cdot(-5\cS_{\text{SU(5)}} + 4  \cS_7 + 4  \cS_9)$\rule{0pt}{1.2Em} \\
 \!\!&\!\!$- \cS_{\text{SU(5)}}\cdot(\cS_7+5 \cS_9) - 2 \cS_7^2 - 
 2 \cS_9^2$\\ \hline
$\mathbf{1}_{(1,1)}$ \!\!&\!\! $6 [K_B^{-1}]^2 + [K_B^{-1}]\cdot (-5\cS_{\text{SU(5)}} - 5 \cS_7  + 4  \cS_9)$\rule{0pt}{1.2Em}\\
 \!\!&\!\!$+ \cS_{\text{SU(5)}} \cdot(3 \cS_7 -5\cS_9)+ \cS_7^2  + 
 \cS_7\cdot \cS_9 - 2 \cS_9^2$ \\ \hline
\end{tabular}
}
\eeq

Finally, we note the presence of  additional hypermultiplets   for a non-simply 
connected GUT-divisor $\cS_{\text{SU}(5)}^b$.
The divisor $\cS_{\text{SU}(5)}^b$ is in general a curve 
of genus $g$ and adds $x_{Adj}=g$ matter hypermultiplets in the adjoint 
representation $\mathbf{25}$ of SU(5) \cite{Katz:1996ht}, which  are 
uncharged under the U(1) gauge fields:
\beq \label{eq:SU5Adj}
\text{
\begin{tabular}{|c|c|}  \hline
Representation & Multiplicity\\\hline
  $\mathbf{24}_{(0,0)}$&  $g$\\ \hline
\end{tabular}
}
\eeq
We note that given an arbitrary divisor $\cS_{\text{SU}(5)}^b$ the genus 
$g$ is calculated via adjunction as
\beq \label{eq:genusSU5}
K_B \cdotp \cS_{\text{SU}(5)}^b + (\cS_{\text{SU}(5)}^b) ^2 =2g-2\,.
\eeq

\subsection{Anomaly cancellation}

The Abelian  anomaly cancellation equations in \eqref{eq:6dAnomalies} can  readily be evaluated  by 
including the non-Abelian representations with their respective 
multiplicities and their number of weights. 
However the curves $b_{mn}$ in the base  are  now modified  by the presence of the $\cS_{\text{SU(5)}}$ divisor, and are  given by
\beq \label{eq:bmnSU5}
b_{mn}= -\pi({\sigma}({\hat s}_n)\cdot {\sigma}({\hat s}_m))=b_{mn}^0  
+(C^{-1})^{ij}(S_m \cdotp {c}_{-\alpha_i})(S_n \cdotp {c}_{-\alpha_j}) \cS_{\text{SU(5)}}\, ,
\eeq
where the Shioda map  $\sigma ({\hat s}_m)$ ($m=1,2$)  is given in 
(\ref{eq:ShiodaMapSQSRSU5}), $b_{mn}^0$ refers to the curves in 
(\ref{eq:b_mnAbelian}), i.e. without  $\cS_{\text{SU(5)}}$,  $(C^{-1})^{ij}$ is 
the inverse Cartan matrix of SU(5) (see (\ref{eq:SU5Cartan})) and  
${c}_{-\alpha_i}$ are the fibral curves corresponding to the simple roots 
of SU(5). After further manipulations,  (\ref{eq:bmnSU5}) can be cast in the following form:
\beq\label{eq:bmnSU5p}
b_{mn}
 = \left( \begin{array}{cc}
                    -2[K_B]-\frac{6}{5}\cS_{\text{SU(5)}} & \cS_9-\cS_7  - [K_B] \\
                    \cS_9-\cS_7  - [K_B] & 2(\cS_9-[K_B])
                   \end{array} \right)_{mn}\, .
\eeq

Comparing the two sides in (\ref{eq:6dAnomalies}) we see that indeed
all Abelian and mixed Abelian-gravitational anomalies are cancelled.
However, in the presence of a non-Abelian gauge sector there are additional anomaly 
equations, see for example \cite{Park:2011wv}. They can be  summarize  as: 
\bea \label{eq:6Danomalies}
\text{Non-Abelian-gravitational anomaly} \ F_\kappa^2R^2:&& \textstyle{ \frac{1}{6}}\left( A_{Adj_\kappa}-\sum_{\mathbf{R}} x_{\mathbf{R}} A_{\mathbf{ R}}\right)=K_B^{-1} \cdotp \left( \frac{b_{\kappa}}{\lambda_\kappa}\right) \,,\nn
\\
\text{Pure non-Abelian anomaly} \ F_\kappa^4: &&\, B_{Adj_\kappa} - \sum_{\mathbf{R}} x_{\mathbf{R}} B_{\mathbf{R}} = 0\,,\nn \\
 \text{Pure non-Abelian anomaly} \ F_\kappa^2F_\kappa^2: && \textstyle{\frac{1}{3}} \left( \sum_{\mathbf{R}} x_{\mathbf{R}} C_{\mathbf{R}}-  C_{Adj_\kappa}  \right) = \left( \frac{b_\kappa}{\lambda_\kappa}\right)^2,\nn \\
 \text{Non-Abelian-Abelian anomaly} \ F_\kappa^3F_{U(1)_i}: &&  \sum_{\mathbf{R},q_i} x_{\mathbf{R},q_i} q_i E_{\mathbf{R}} = 0\,,\nn \\
  \text{Non-Abelian-Abelian anomaly} \ F_\kappa^2F_{U(1)_i}F_{U(1)_j}: && \sum_{\mathbf{R},q_i,q_j} x_{\mathbf{R},q_i,q_j} q_i q_j A_{\mathbf{R}} = \left( \frac{b_\kappa}{\lambda_{\kappa}} \right) \cdotp b_{ij}\,.\nn \\
\eea
Here, the `$x_{\cdot}$'  variables denote the multiplicities of 
hypermultiplets: $x_{\mathbf{R}}$,  $x_{\mathbf{R},q_i}$ and $x_{\mathbf{R},q_i,q_j}$   are the number of hypermultiplets in 
the $\mathbf{R}$ representation, in the 
$\mathbf{R}$ representation with charge $q_i$ under U(1)$_i$ and 
in the $\mathbf{R}$ representation with 
charges $\{q_i,q_j\}$ under U(1)$_i
\times$U(1)$_j$, respectively.
Furthermore,  we used 
the following group theory relations:
\beq \label{eq:groupfactors}
\text{tr}_{\mathbf{R}} F_\kappa ^2=A_{\mathbf{R}}\text{tr} F_\kappa ^2\,,\qquad \text{tr}_{\mathbf{R}} F_\kappa ^3=E_{\mathbf{R}}\text{tr} F_\kappa ^3\,,\qquad\text{tr}_{\mathbf{R}} F_\kappa ^4=B_{\mathbf{R}}\text{tr} F_\kappa ^4+C_{\mathbf{R}}(\text{tr} F_\kappa ^2)^2\,.
\eeq
Here `tr' denotes the trace with respect to the fundamental representation 
and $\text{tr}_{\mathbf{R}}$ is the trace for a given  representation ${\mathbf{R}}$. 
For $\kappa=\text{SU}(N)$,   the group theory factors (\ref{eq:groupfactors}) have the  following
values:\footnote{See \cite{Taylor:2011wt} for further details.}
\beq
\text{
\begin{tabular}{|c|c|c|c|c|c|} \hline
 Representation & Dimension & $A_{\mathbf{R}}$ & $B_{\mathbf{R}}$ & $C_{\mathbf{R}}$ & $E_{\mathbf{R}}$  \\ \hline
 Fundamental & $N$ &1 & 1& 0& 1 \rule{0pt}{1Em}\\ \hline
 Adjoint & $N^2-1$ &$2N$ & $2N$  & 6 & $2N$\rule{0pt}{1Em} \\ \hline
 Antisymmetric& $N(N-1)/2$ & $N-2$ & $N-8$ & 3 & 1 \rule{0pt}{1Em}\\ \hline
\end{tabular}
}
\eeq
 Furthermore,   $\lambda_\kappa$ in  (\ref{eq:6Danomalies})  is the group 
 normalization constant defined by 
 $\lambda_\kappa=2c_{G(I)}/E_{adj_\kappa}$, where $c_{G(I)}$ is the dual 
 Coxeter number and $E_{\mathbf{R}}$ is defined in  
 (\ref{eq:groupfactors}). 
For $\kappa=\text{SU}(N)$,  $\lambda_{\kappa}=1$.   Finally, $b_\kappa$  
in (\ref{eq:6Danomalies}) is the divisor supporting the non-Abelian group 
in the base; in our case    $\kappa=\text{SU}(5)$ and $b_\kappa = 
\cS_{\text{SU}(5)}$.

With the above information we can evaluate explicitly the anomaly  cancellation equations  (\ref{eq:6Danomalies}).
In our case we have to sum over the fundamental, 
anti-symmetric and adjoint representations,  along with their 
multiplicities  (\ref{eq:mults5}) and 
(\ref{eq:SU5Adj}), and   the expression (\ref{eq:bmnSU5p})  for $b_{mn}$.   The upshot is  that all the anomaly cancellation equations
(\ref{eq:6Danomalies}) are satisfied. Thus, the generic F-theory compactification 
with the chosen SU(5)$\times$U(1)$\times$U(1) gauge symmetry over an
arbitrary base $B$ is anomaly free.

\subsubsection*{Acknowledgments}

 We would like to thank Thomas Grimm, Jim Halverson, Albrecht Klemm, 
 Sakura Sch\"afer-Nameki, Peng Song and Washington Taylor for 
 discussions.  We are  especially indebted to Antonella Grassi for many 
 useful discussions and comments. M.C.~is grateful to  the Theory 
 Division of CERN for 
hospitality. D.K. thanks the University of Hamburg and DESY for 
hospitality. H.P is grateful to PiTP  at the Institute for Advanced Study 
for hospitality. This research is supported in part by
the DOE grant DE-SC0007901 (M.C., H.P., D.K.), the NSF String Vacuum 
Project Grant No. NSF PHY05-51164 (H.P.), Dean's Funds for Faculty 
Working Group (M.C. and D.K.), the Fay R. and Eugene L. Langberg Endowed 
Chair (M.C.) and the Slovenian Research Agency
(ARRS) (M.C.).

\bibliographystyle{utphys}	% (uses file "plain.bst")
\bibliography{ref}

\end{document}